\documentclass[11pt]{article}
\usepackage{hyperref}
\usepackage{graphicx}
\usepackage{fullpage}

\begin{document}

\begin{center}
{\huge {\bf 
HTC Scientific Computing in a \\
Distributed Cloud Environment}}
\end{center}

\vspace{1cm}

\begin{center}
R.J. Sobie, A. Agarwal, I. Gable, 
C. Leavett-Brown, M. Paterson, R. Taylor \\
{\it Institute of Particle Physics} \\
{\it University of Victoria} \\
{\it Victoria, Canada}
\end{center}

\begin{center}
A. Charbonneau, R. Impey, W. Podiama \\
{\it National Research Council of Canada} \\
{\it Ottawa, Canada }
\end{center}

\vspace{1cm}


\begin{center}
{\large\bf Abstract }
\end{center}

This paper describes the use of a distributed cloud computing system 
for high-throughput computing (HTC) scientific applications.
The distributed cloud computing system is composed of a number of separate 
Infrastructure-as-a-Service (IaaS) clouds that are utilized in a unified 
infrastructure. 
The distributed cloud has been in production-quality operation for two years 
with approximately 500,000 completed jobs where a typical 
workload has 500 simultaneous embarrassingly-parallel jobs that run for 
approximately 12 hours.
We review the design and implementation of the system which is 
based on pre-existing components and a number of custom components.
We discuss the operation of the system, and describe our plans for the 
expansion to more sites and increased computing capacity.

\newpage
\section{Introduction}
Large scientific projects are increasingly becoming international 
in scope with each national group providing computational resources 
to the collaboration.
Computational grids are designed to unify the distributed resources into 
a single infrastructure \cite{foster}; and the WLCG \cite{wlcg}, used by the 
experiments at the Large Hadron Collider project,
is an example of a successfully operating computational grid.
One of the challenges of a computational grid is that the use of
the remote resources is subject to the local policies and software 
implemented at each site.  
Virtualization technologies offer a mechanism for minimizing the
dependence on the local configuration by encapsulating the 
complex application software suite into a virtual machine (VM) image.
It has been shown that particle physics applications, for example, 
can run in a VM image without any loss in efficiency \cite{gablebenchmarks}.
There are many examples of scientific projects running large numbers
of jobs on Infrastructure-as-a-Service (IaaS)
academic clouds\footnote{An academic
cloud is one that is operated by a research organization such as a
university or laboratory on behalf of its constituency} 
or commercial clouds (for example, see \cite{magellan}, 
\cite{starcloudcomputing} or \cite{bellecloudcomputing}).

Recently it was pointed out that a ``grid of clouds" or ``sky computing" 
is a potential solution for combining separate IaaS clouds into a 
unified infrastructure \cite{skycomputing}.
We have been operating a distributed cloud computing system
for a number of years for the particle physics and astronomy 
communities in Canada.
The computational resources in Canada, like in most countries, are
distributed and shared with many user communities.
A distributed cloud computing system provides important benefits:
virtualization shields the application suite from changing
technologies and reduces the need for systems personnel to be 
knowledgeable about the user application;
IaaS clouds provide a simple way to dynamically manage the load
between multiple projects within a single center; and a distributed 
cloud aggregates heterogeneous clouds into a unified resource with 
a single end-point for the users.

The design and initial operation of our distributed cloud computing 
system has been discussed in an earlier publication \cite{hpcs:cloudpaper}.
Over the past year the system has been expanded to include more IaaS 
clouds and applications.
We are currently running an average of 500 simultaneous jobs  
with peaks of approximately 1000 simultaneous jobs.
We estimate that nearly 500,000 particle physics and
astronomy jobs have been completed.
We continue to improve the capabilities of the system.
Recently we have added the ability to simultaneously use multiple types of 
IaaS clouds and  support multiple hypervisors.
We have enhanced the  monitoring and management tools, 
and enabled the system to use remote software and data repositories.
We review the design of the distributed cloud computing system
and discuss the results using particle physics applications.   
We describe our plans to increase the computing capabilities of the 
distributed cloud computing system and run data intensive applications.

\section{Design overview}

We give an overview of the components used in the distributed 
cloud computing system.
We discuss the interactive cloud service, user authentication and 
delegation, VM image repository, batch cloud services, software 
repository, data repository, and system monitoring and 
management tools. 

\subsection*{\it Interactive cloud services}

An interactive service is provided with a single head node configured 
with Scientific Linux and a set of VM image management tools.
The user has access to a VM image repository which can be used to 
store customized VM images (described below).

Users submit batch jobs from the interactive system using HTCondor 
\cite{condor} as if they were submitting jobs to a conventional 
HTC batch computing system.
The user can query the batch queue, cancel jobs and retrieve the output
of their jobs using the HTCondor commands.
Once a user submits their jobs, the instantiation of the VM image on 
the clouds and the assignment of jobs to the available resources 
is hidden from the user.
However, the user can specify to run their job on a single cloud or 
set of clouds by setting a custom configuration option in the HTCondor 
job description file.

A full description of our interactive system is found in Ref.~\cite{colinpaper}.
Our astronomy colleagues provide a similar but separate interactive service.

\subsection*{\it User authentication and delegation}

We use X.509 certificates \cite{x509} to authenticate users, and the 
MyProxy Credential Management System \cite{myproxy} to obtain proxy 
credentials that are used to delegate the user's authority to the
system services.

The user uploads their credentials to a MyProxy server and creates 
a 12-hour proxy certificate from the MyProxy server.
The proxy certificate is used to authenticate the user's interaction
with the VM image repository, the batch submission system and cloud
scheduling service.

A user's credentials must be authenticated and authorized before they can 
boot or stop VM images on a Nimubus cloud  \cite{nimbus}.
A group authentication, using a single shared access key, is used 
on OpenStack \cite{openstack} clouds.

\subsection*{\it Virtual machine image repository}

We have developed a VM image repository, called Repoman \cite{repoman}, 
that is designed to operate as a standalone system and be independent 
of the type of IaaS cloud.
At the beginning of the project, there were no VM image repositories that
met our requirements, however, today there are alternative VM image 
repositories such as Glance in OpenStack \cite{glance} which provides 
VM image management on OpenStack installations; and the FutureGrid Image 
Repository \cite{futuregrid-repo} which offers similar functionality to Repoman.

Repoman is implemented as a RESTful web service authenticated with 
X509 certificates as discussed in the previous section.
The user can boot a VM image, customize and save the 
image in the Repoman repository.
VM images are accessible via HTTP(S) and retrieved with tools such 
as wget and curl. 
A detailed description of Repoman is found in ref.~\cite{repoman}.

A new feature of Repoman, not previously described, is its ability 
to manage VM images capable of running under both the KVM \cite{kvm}
and Xen \cite{xen} hypervisors. 
Such an image is said to be a ``dual-hypervisor" image. 
The instance creation services of both the batch and interactive 
systems use this new feature to start dual-hypervisor VM images seamlessly 
on any of the available clouds, regardless of hypervisor type (KVM or Xen). 
Neither the OpenStack Glance nor FutureGrid Image Repository 
provide this feature.

\begin{figure}[ht]
\begin{center}
\includegraphics[height=10cm]{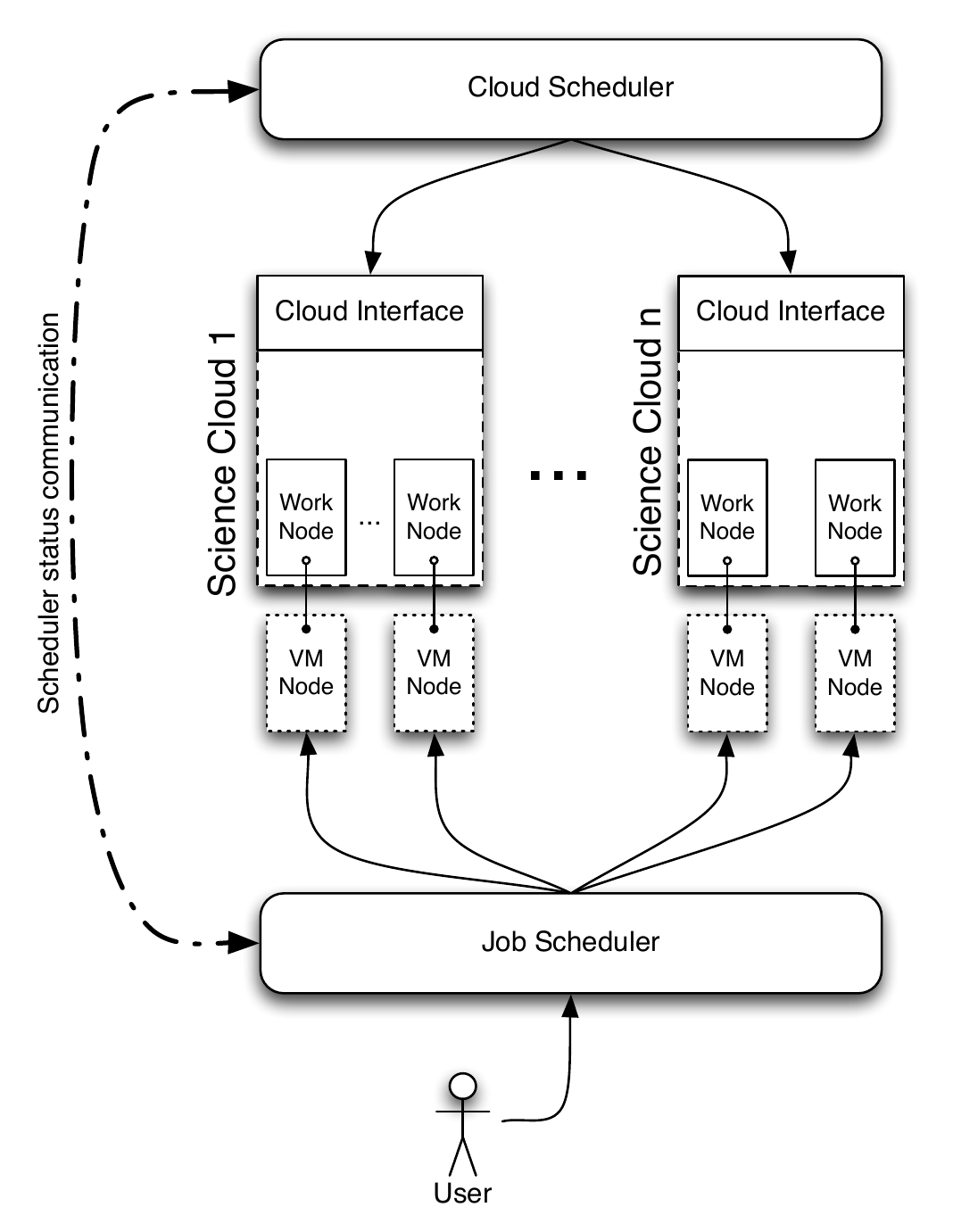}
\end{center}
\caption{\label{fig1} 
An overview of the architecture used for the system.
A user prepares their VM image and a job script.  
The job script is submitted to the HTCondor job scheduler.
Cloud Scheduler reads the job queue and makes a request to boot the 
user VM on one of the available clouds.
Once the VM is booted, it attaches itself to the HTCondor pool and
HTCondor assigns jobs to the VM image.
When there are no more user jobs requiring that VM type, Cloud Scheduler 
makes a request to the proper cloud to shutdown the user VM.
}
\end{figure}

\subsection*{\it Batch cloud services}

The scheduling of jobs is performed by HTCondor \cite{condor} and 
the deployment of VM images is done by Cloud Scheduler \cite{cloudscheduler}
(see fig.~\ref{fig1}).

HTCondor is a well-known job scheduler similar to PBS \cite{pbs} and
GridEngine \cite{gridengine}. 
HTCondor was designed as a cycle scavenger making it an ideal job 
scheduler for a dynamic environment where resources (VM instances) 
appear and disappear based on demand.
HTCondor manages a queue of jobs that are ordered by the submission time.
Periodically it cycles through the queue and submits the 
highest priority job\footnote{Apart from the ``Priority-FIFO" policy, 
HTCondor also supports both job and user priorities. For more details see 
\url{http://research.cs.wisc.edu/htcondor/manual/v7.8/2\_7Priorities\_Preemption.html}.}
when the job requirements can be matched with the properties of a resource.
We have selected a simple scheduling algorithm as the system is 
currently used by a limited number of users.

The user prepares an HTCondor job description file that specifies the
VM image required for the job and the system requirements for 
instantiating the VM image on the clouds.

For Nimbus clouds, the user must provide a URL to the location of the 
VM image, the path to the user's proxy credential, the host name of 
the MyProxy server, the system architecture, the number of cores 
and the memory size.
In addition, the user can include temporary storage by requesting 
Nimbus attach an optional block device, known as a blank space partition,
of a specified size.
We have observed that Nimbus does not provide any safeguards against 
over allocation of the temporary storage which can result in I/O errors 
to the block device. 
This issue is mitigated if the cloud provides sufficient resources to satisfy 
the storage requests.

The user must give the AMI (Amazon Machine Image) identifier of the 
VM image and instance type for OpenStack clouds.
The instance type determines the parameters of the various VM images
(e.g. CPU, storage and memory) that are available to users.

Once the job description file is complete, then the user issues the 
{\it condor\_submit} command to add the job to the HTCondor job queue.
Alternatively, if the job is part of a collection of inter-dependent tasks,  
the user can include the job in a Directed Acyclic Graph (DAG) input file 
to be submitted by the {\it condor\_submit\_dag} command \cite{condor}. 
The HTCondor DAG Manager manages the sequence and conditional execution 
of all jobs within a DAG input file, and is completely compatible with 
the regular HTCondor job queue and the operations of Cloud Scheduler.

The management of the VMs in the distributed cloud computing system is 
done by a customized component called Cloud Scheduler \cite{cloudscheduler}.
HTCondor recently added the ability to boot VM images \cite{condor}, 
however, the implementation necessitates that the user manage their 
resources as well as their jobs.
Cloud Scheduler periodically reviews the requirements of the jobs in the 
HTCondor job queue and makes requests to boot user-specific VM images on 
one of the IaaS clouds (see fig.~\ref{fig1}).
Once the VM image is booted, then the condor worker starts running and adds 
itself to the HTCondor resource pool, and is ready to accept jobs.
Cloud Scheduler selects the user job with the highest priority in the 
HTCondor job queue and searches for a cloud where the user VM can be booted.
Once a VM image is booted for the first user, Cloud Scheduler re-examines 
the job queue for the highest priority job of a different user and requests 
the instantiation of this user's VM image on one of the clouds.
This procedure is repeated for each user in the job queue until all users have 
an opportunity to start one job in a Cloud Scheduler scheduling cycle.
If there are no available resources for a job of a given user, then 
Cloud Scheduler searches for another job from the same user with a 
different VM type before looking at the next user.

The job and VM scheduling algorithm is based on our operational experience.
We anticipate that the scheduling algorithms will evolve as the system
scales to more users and resources.
For example, we have observed that the scheduling of VM images 
is not fully efficient when there are user requests for both
single-core and whole-node VM images.
We are considering a simple solution that would separate the jobs 
requiring single-core and whole-node VM images.

Cloud Scheduler is able to load-balance the distributed cloud by 
equalizing the number of instantiated VM images of the users.
Currently we have configured Cloud Scheduler to re-balance the system 
by retiring one VM instance of the user with the most number of 
instantiated VMs and then request that a new VM of the next user be
booted on a cloud.
Cloud Scheduler retires an existing VM instance by sending the HTCondor 
server a request to issue a {\it condor\_off} command to shut down the 
HTCondor daemons on the existing VM instance. 
This will stop new jobs from starting on the existing VM instance and let 
the current job finish.

Cloud Scheduler will shutdown VM instances in an error state 
(as reported by the cloud platform).
VM instances can also be stopped by the local system (e.g. the
IaaS cloud is down for maintenance).
If the resource (VM instance) used by a job disappears, then HTCondor
reschedules the job for execution and Cloud Scheduler will
request a new VM image be booted on one of the clouds.

It is important that Cloud Scheduler maintain a valid proxy certificate
for each user otherwise it loses the ability to manage the user's 
VM instances on Nimbus clouds. 
To avoid the system going into such a state, Cloud Scheduler will 
shutdown a VM instance on a Nimbus cloud if the validity of the 
user's proxy certificate is about to expire. 

VM instances on a Nimbus cloud have a default lifetime of seven days
whereas OpenStack clouds have an infinite lifetime.
Cloud Scheduler tracks the lifetime of the VM instances and will shut the 
instance down before the end of its life.

\begin{figure}[ht]
\begin{center}
\includegraphics[width=12cm]{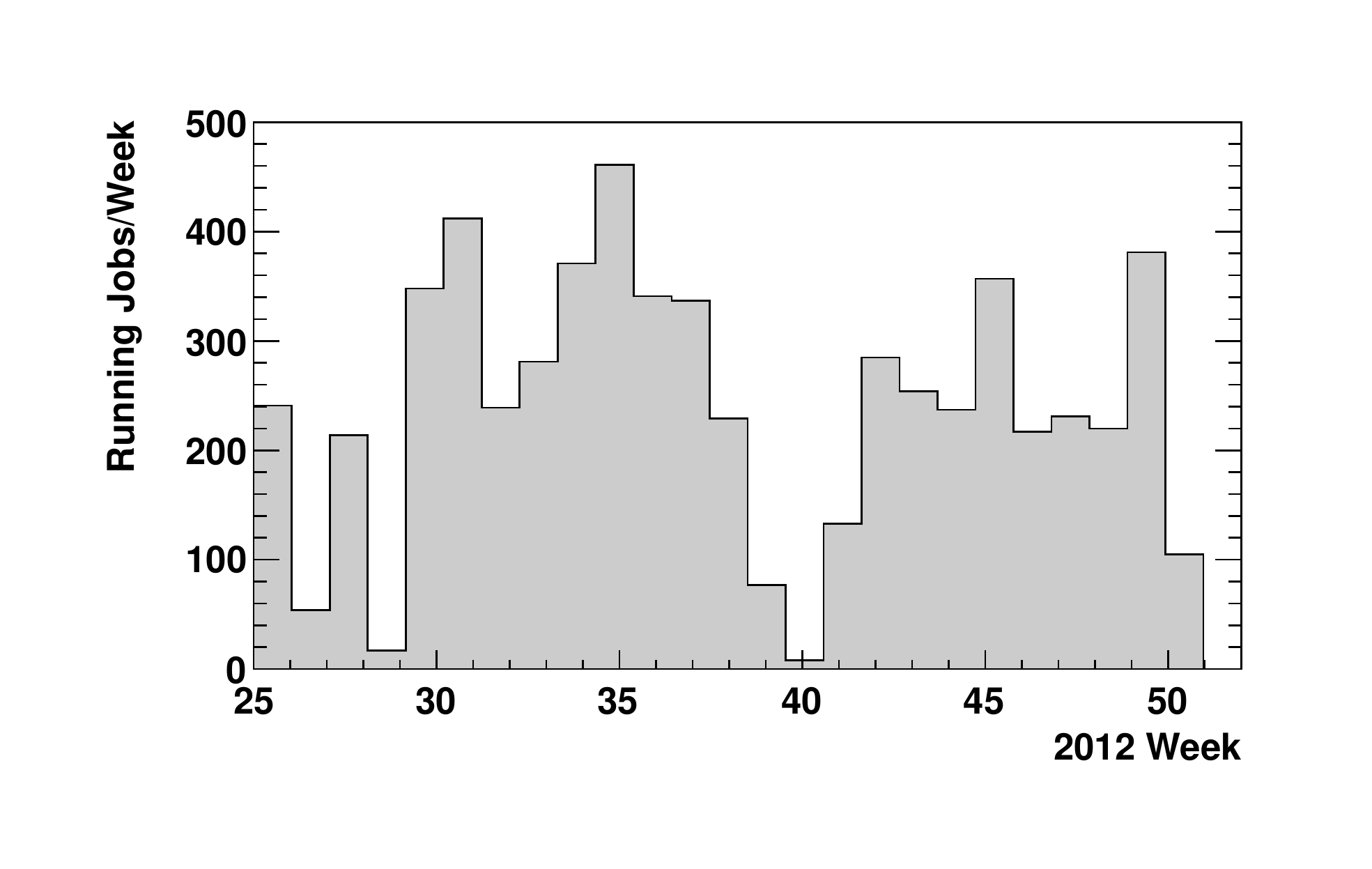}
\end{center}
\caption{\label{fig:runningjobs} 
The average number of simultaneously running particle physics jobs 
for each week between July 1 and December 31 2012.
The label on the horizontal axis is the week number of the year.
The average number of jobs is approximately 250 with peaks 
approaching 500 jobs.
The Melbourne cloud joined the system in December and is not
included in this plot.
}
\end{figure}

\subsection*{\it Software repositories}

We use the CernVM file system (CVMFS) \cite{cvmfs} for managing 
application software in a virtual environment.
Each suite of application software is stored in a separate appliance,
which presents a POSIX file system tree to the clients.
The VM images do not have a copy of the application software
but use the CVMFS client, employing the HTTP protocol and FUSE file system, 
to view the complete application tree and stage-in the required components.
CVMFS reduces the size of the VM image and minimizes the time to propagate 
the image to distant clouds as the required software files will only be uploaded 
as they are requested.

\subsection*{\it Data repository}

We provide a read-only data repository to the batch VM instances 
with write access available through the interactive system.
We use a two-tier approach for the data repository, with 
Lustre \cite{lustre} as the filesystem for backend data storage and  
Apache2/WebDAV \cite{webdav} for the frontend data distribution.

Lustre provides a single name space, large capacity, and load sharing 
over multiple servers.
Apache2/WebDAV uses HTTP(S) protocols to present a POSIX compliant filesystem.
The WebDAV client, like CVMFS, presents a view of the entire filesystem 
tree but only stages the file content as it is accessed.
WebDAV supports both read and write requests from clients; however, 
we restrict clients to read-only access to the repository by using the
Apache2  ``LimitExcept" directives.
Our preliminary studies indicate that WebDAV provides superior 
performance compared to the other data repository access methods. 

The output data from the batch jobs is written to ephemeral storage 
attached to the VM instance.
HTCondor commands are employed by the user on the interactive system to
retrieve the data. 
If any output data is produced in an interactive VM instance, then the user 
is responsible for transferring their data to permanent storage using 
data transfer tools typically used in a grid environment (e.g. gridFTP).

\subsection*{\it System monitoring and management tools}

We have developed a set of management tools for the administrator.
The tools allow the administrator to dynamically add or remove IaaS clouds
from the batch system.
The administrator has the ability to stop running VM instances if necessary.
Users have access to tools on the interactive system provided by 
HTCondor and Repoman as well as a number of customized tools.

We monitor the number of VM instances and jobs on the entire system and 
on each cloud.
All the servers provide detailed monitoring information that can 
be graphically displayed or written to a file for later analysis.
If the VM image includes a monitoring option, then detailed information
on each VM instance is recorded and available for display.

\begin{figure}[ht]
\begin{center}
\includegraphics[width=10cm]{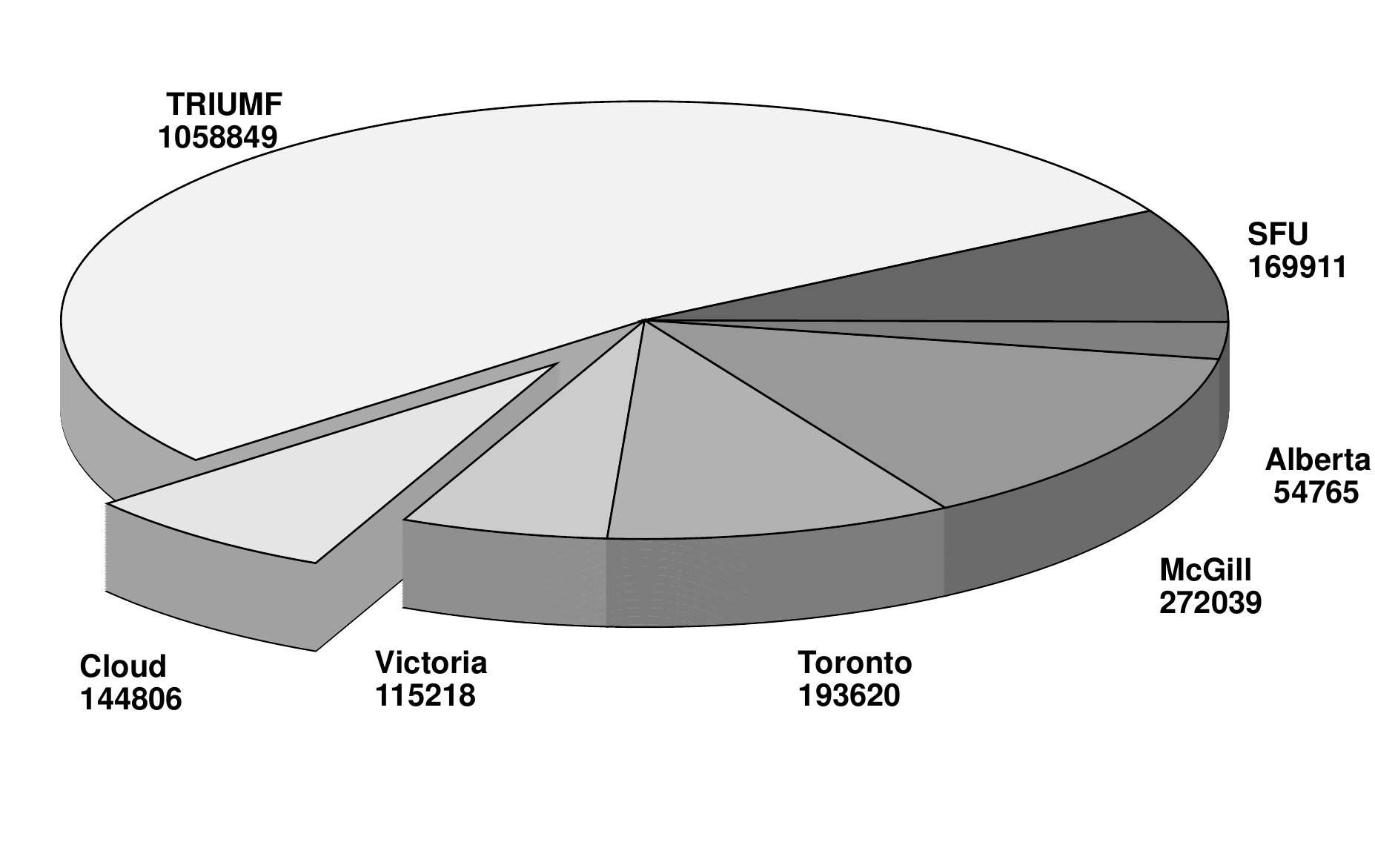}
\end{center}
\caption{\label{fig:piechart} 
The pie chart shows the number of completed simulation production jobs 
on the Canadian resources used by ATLAS between July 1 and December 31 2012.
There is a Tier-1 facility in TRIUMF, and five Tier-2 
facilities in Victoria, Toronto, Montreal, Edmonton and Vancouver.
The distributed cloud ran approximately 7\% of the simulation jobs;
however, the other centers also run the analysis jobs which are not
included in the total number of jobs.
}
\end{figure}

\begin{figure}[ht]
\begin{center}
\includegraphics[width=8cm]{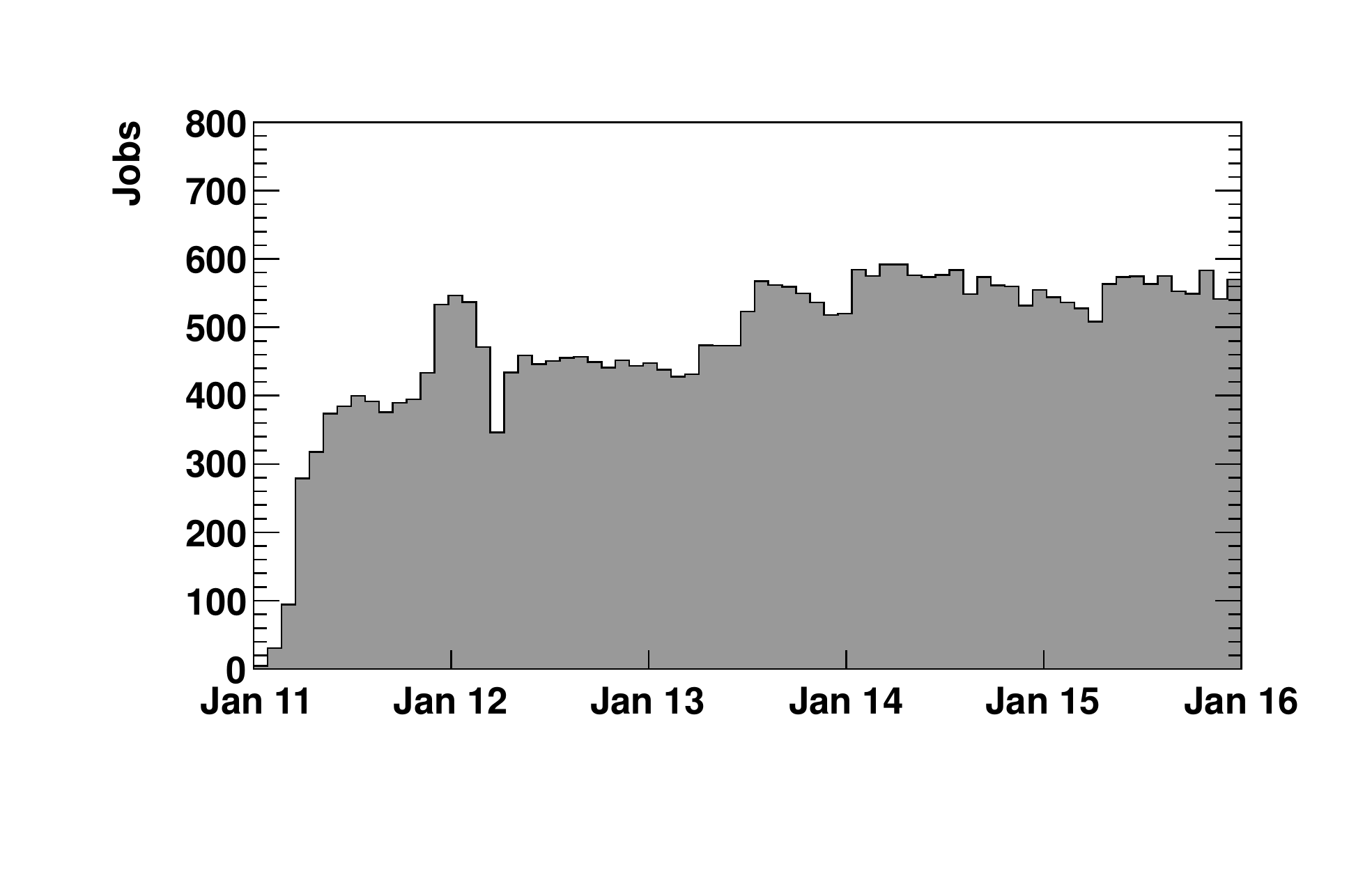}
\includegraphics[width=8cm]{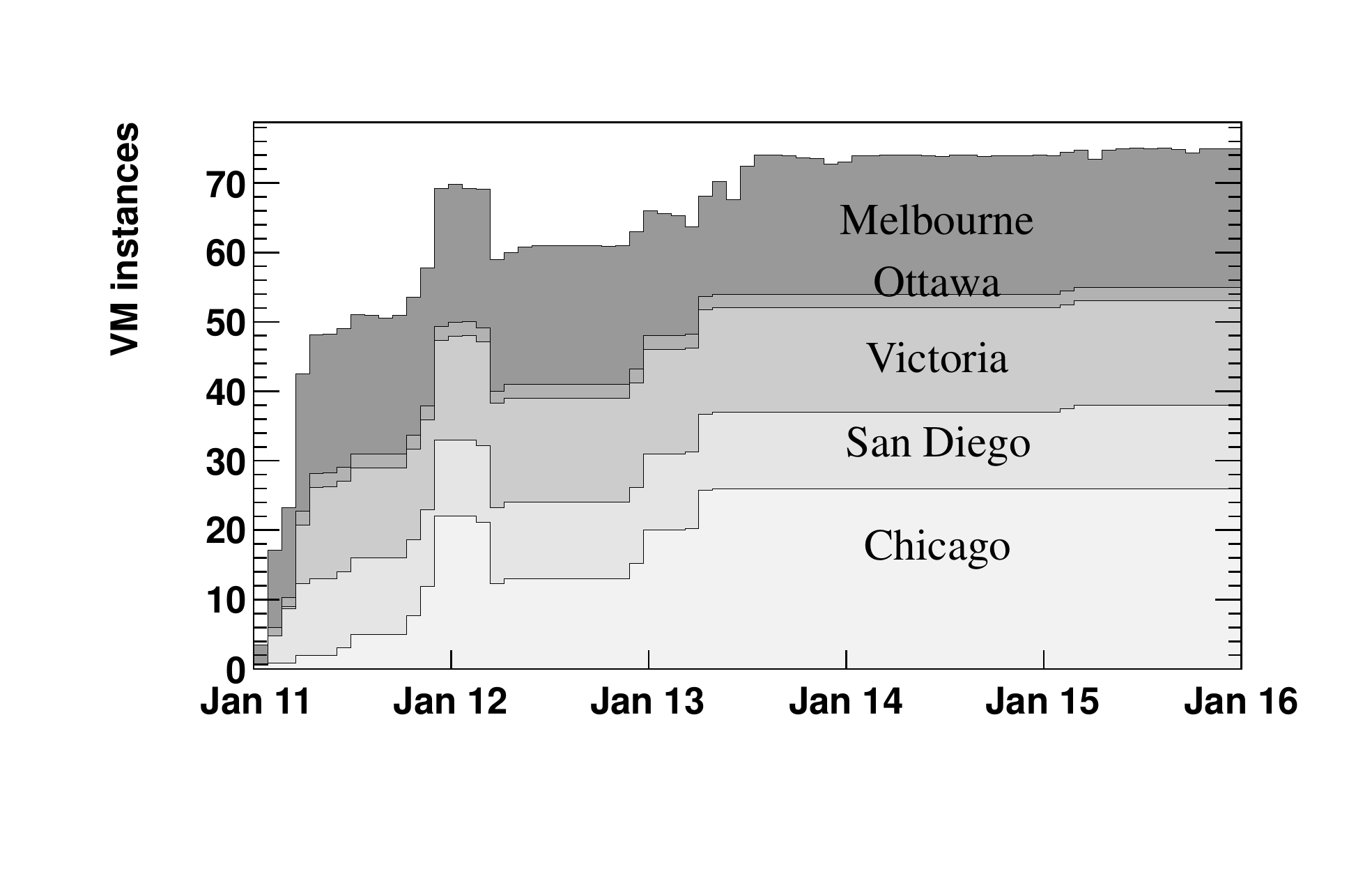}
\end{center}
\caption{\label{fig2} 
The number of jobs and VM instances on the distributed cloud 
computing system for a five day period starting January 11 2013.
The plot of VM instances includes five separate IaaS clouds. 
All the VM instances in this plot are whole-node VM instances.
}
\end{figure}

\section{Results}

This section gives a description of the applications
used in the distributed cloud computing system and
a discussion of the system performance.

\subsection*{\it Application image}

The particle physics applications used on the distributed computing 
cloud system are from the BaBar experiment based 
at the SLAC National Laboratory Center in Palo Alto \cite{babar} and 
the ATLAS experiment based at the CERN Laboratory in Geneva \cite{atlas}.
The system has been used to run production jobs from both 
collaborations as well as local user analysis jobs.
Production jobs are centrally controlled by a small team and are used 
to reconstruct the raw data into the relevant quantities for physics
analysis or to generate simulated data (Monte Carlo) samples which
are used to develop selection algorithms and understand the response
of the detector systems.

We have described the use of the distributed cloud for the BaBar
experiment in a previous publication \cite{colinpaper}.
The BaBar experiment stopped recording particle collision data in 2008 
and its current demands for production jobs using external resources is small.
Currently we use the distributed cloud for local analysis of BaBar data samples.

We have created a VM image for the ATLAS production jobs and a 
second VM image for the ATLAS and BaBar analysis jobs.
The production-VM image uses CernVM 2.6.0 \cite{cernvm}, 
Scientific Linux 5.8, and three CVMFS repositories provided by CERN
(ATLAS software with 23 million files and 1.3 TB;
ATLAS calibration data with 8000 files and 473 GB; and
grid utility software with 200,000 files and 3.4 GB).
The size of the VM image is approximately 9 GB.
and is instantiated with a 100 GB ephemeral block device mounted 
at /tmp for all temporary input and output data.

The analysis-VM image uses Scientific Linux 5.7 with an additional 
BaBar repository (200,000 files and 11 GB).
The image size is 3.8 GB and is also instantiated with 100 GB 
of storage.
The production-VM image (CernVM) is only 28\% full with the rest of 
the space used for caching and swap space whereas the analysis-VM image
has only a small amount of cache space and no swap space in the image. 
The typical time required to save a VM image in Repoman is
approximately 1 minute per Gigabyte.

There are significant benefits if the VM instance uses all the 
cores of the node (``whole-node'' VM instances).
Whole-node VM instances reduce the number of VM images that need to
be transferred from the Repoman image repository, making efficient 
network use and minimizing the load on the VM image servers.
Whole-node VM instances require only one CVMFS cache per physical node 
for all the jobs, making more efficient use of the local storage and 
eliminating duplicate transfers to the cache and disk contention.

\subsection*{\it System performance}

The distributed cloud computing system uses IaaS academic clouds in 
Canada (Victoria and Ottawa), the United States (FutureGrid clouds in 
San Diego and Chicago) \cite{futuregrid} and Australia (Melbourne).
In Victoria, we have two IaaS clouds with one belonging to the 
particle physics group and another owned by Compute Canada
\cite{computecanada}.
The Victoria and FutureGrid clouds run Nimbus while the 
Melbourne cloud uses OpenStack.
One of the Victoria clouds and the FutureGrid clouds use Xen images
while the second Victoria cloud and Melbourne use KVM images.
Our astronomy colleagues use a different set of clouds for their
applications, however, we share the Victoria Compute Canada cloud.
We are testing a number of commercial clouds; however, 
we find that the commercial clouds often do not provide the type of image 
we require for our applications (e.g. particle physics applications 
typically use older releases of the Linux operating system)
and in other cases our use is limited as our connection is via 
a low-bandwidth commercial network.

The distributed cloud computing system is integrated with the Victoria 
ATLAS Tier-2 center providing an extension of the compute resources.
A Tier-2 center is used to generate simulated data samples and 
provide computational resources for the individual researchers.
Our system is currently restricted to applications with relatively 
low I/O data requirements.   
Further study is required before higher I/O production jobs can be run 
on the system.

We show the average number of simultaneously running ATLAS jobs per week 
on the system for the six month period from July to December 2012
in fig.~\ref{fig:runningjobs}.
The plot does not include the jobs from the Melbourne cloud which joined
the system in December 2012.
Typically, 250 jobs were running in this period with peaks of 500 
simultaneous jobs.
Fig.~\ref{fig:piechart} shows that there were approximately 144,000
completed ATLAS jobs on the distributed cloud computing system, accounting
for about 7\% of the total Canadian simulation production\footnote{
Note that the other Canadian sites also run analysis jobs in addition
to the simulation production jobs.} in the same period.
The processing efficiency of the simulation production jobs on the
Canadian resources is 97.4- 98.7\% and 93.0\% for the cloud resources
where the efficiency is defined to be the CPU time divided by the 
wallclock time.
The remote resources must retrieve a VM image and access the CVMFS
server in Victoria.
We have made no attempt to optimize the processing efficiency.

In the top plot of fig.~\ref{fig2}, we show the number of ATLAS 
production jobs running in a five-day period starting January 11 2013.
The bottom plot shows the number and location of VM instances during
the same period.
The IaaS clouds active at this time were Melbourne, Ottawa, Victoria, 
San Diego and Chicago.  
Each VM instance is a whole-node (8-core) instance.
We observe that the number of jobs fluctuates due to a number of factors:
the ATLAS jobs are automatically assigned to our system by a central 
production team; and the individual clouds have the ability to modify 
the number of VM instances on their system based on their local demand 
and local maintenance schedules.

\section{Summary}
We have described the design and operation of a distributed cloud 
computing system for high-throughput computing (HTC) scientific applications.
The system is composed of a number of separate IaaS clouds that are 
utilized in a unified infrastructure for applications in particle 
physics and astronomy.
The distributed cloud has been in production-quality operation for two years 
with approximately 500,000 completed jobs.

The short-term plans are to scale the distributed cloud computing system 
to run more than 1000 simultaneous jobs by adding three new IaaS academic 
clouds in 2013.
Currently, the system has been primarily used for low I/O applications, 
however, we are able to run high I/O particle physics applications on 
the two clouds in Victoria, which are co-located with the data repository.
However, running more high I/O jobs on the full system will require a more 
sophisticated approach to managing the data in a distributed fashion.
The WLCG project \cite{wlcg} has devoted a significant amount of 
effort studying the best techniques for managing data in a wide-area
computing grid and we plan to leverage their developments.

\section*{Acknowledgments}
The support of CANARIE, the Natural Sciences and Engineering Research Council, 
the National Research Council of Canada, and FutureGrid are acknowledged.
FutureGrid is supported in part by the National Science Foundation under 
Grant No. 0910812.

\bibliographystyle{abbrv}
\bibliography{apaper}

\begin{thebibliography}{10}

\bibitem{colinpaper}
A.~Agarwal, M.~Anderson, P.~Armstrong, A.~Charbonneau, R.Demarais, K.~Fransham,
  I.~Gable, D.~Harris, R.~Impey, C.~Leavett-Brown, M.~Paterson, W.~Podaima,
  R.~Sobie, and M.~Vliet.
\newblock Simulation and user analysis of {BaBar} data in a distributed cloud.
\newblock {\em PoS}, ISGC 2011 and OGF 31:086, 2011.

\bibitem{gablebenchmarks}
M.~Alef and I.~Gable.
\newblock {HEP} specific benchmarks of virtual machines on multi-core {CPU}
  architectures.
\newblock {\em Journal of Physics: Conference Series}, 219(5):052015, 2010.

\bibitem{cloudscheduler}
P.~Armstrong, A.~Agarwal, A.~Bishop, A.~Charbonneau, R.~Desmarais, K.~Fransham,
  N.~Hill, I.Gable, S.Gaudet, S.Goliath, R.Impey, C.~Leavett-Brown,
  J.~Ouellete, M.~Paterson, C.~Pritchet, D.~Penfold-Brown, W.~Podaima,
  D.~Schade, and R.~Sobie.
\newblock Cloud {S}cheduler: a resource manager for a distributed compute
  cloud.
\newblock arXiv:1007.0050v1 [cs.DC].

\bibitem{atlas}
{ATLAS} experiment at the {CERN} {L}aboratory in {G}eneva.
\newblock \url{http://www.cern.ch/}.

\bibitem{babar}
{BABAR} experiment at the {SLAC} {N}ational {A}ccelerator {L}aboratory in
  {P}alo {A}lto.
\newblock \url{http://www-public.slac.stanford.edu/babar/}.

\bibitem{starcloudcomputing}
J.~Balewski, J.~Lauret, D.~Olson, I.~Sakrejda, D.~Arkhipkin, J.~Bresnahan,
  K.~Keahey, J.~Porter, J.~Stevens, and M.~Walker.
\newblock Offloading peak processing to virtual farm by {STAR} experiment at
  {RHIC}.
\newblock {\em Journal of Physics: Conference Series}, 368(1):012011, 2012.

\bibitem{xen}
P.~Barham, B.~Dragovic, K.~Fraser, S.~Hand, T.~Harris, A.~Ho, R.~Neugebauer,
  I.~Pratt, and A.~Warfield.
\newblock Xen and the art of virtualization.
\newblock In {\em Proceedings of the 19th {ACM} {S}ymposium on {O}perating
  {S}ystems principles}, SOSP '03, pages 164--177, New York, 2003.

\bibitem{wlcg}
I.~Bird.
\newblock Computing for the {L}arge {H}adron {C}ollider.
\newblock {\em Annual Review of Nuclear and Particle Science}, 61(1):99--118,
  2011.

\bibitem{cvmfs}
J.~Blomer, P.~Buncic, and T.~Fuhrmann.
\newblock {CernVM-FS}: delivering scientific software to globally distributed
  computing resources.
\newblock In {\em Proceedings of the first international workshop on
  Network-aware data management}, NDM '11, pages 49--56, New York, NY, USA,
  2011. ACM.

\bibitem{cernvm}
P.~Buncic, C.~A. Sanchez, J.~Blomer, L.~Franco, A.~Harutyunian, P.~Mato, and
  Y.~Yao.
\newblock {CernVM -} a virtual software appliance for lhc applications.
\newblock {\em Journal of Physics: Conference Series}, 219(4):042003, 2010.

\bibitem{hpcs:cloudpaper}
A.~Charbonneau, A.~Agarwal, M.~Anderson, P.~Armstrong, K.~Fransham, I.~Gable,
  D.~Harris, R.~Impey, C.~Leavett-Brown, M.~Paterson, W.~Podaima, R.~Sobie, and
  M.~Vliet.
\newblock Data intensive high energy physics analysis in a distributed cloud.
\newblock {\em Journal of Physics: Conference Series}, 341:012003, 2012.

\bibitem{computecanada}
{C}ompute {C}anada.
\newblock \url{http://computecanada.ca/}.

\bibitem{futuregrid-repo}
J.~Diaz, G.~von Laszewski, F.~Wang, A.~J. Younge, and G.~C. Fox.
\newblock Futuregrid image repository: A generic catalog and storage system for
  heterogeneous virtual machine images.
\newblock In {\em Third IEEE International Conference on Coud Computing
  Technology and Science (CloudCom2011)}, Athens, Greece, 2011.

\bibitem{myproxy}
T.~Fleury, J.~Basney, and V.~Welch.
\newblock Single sign-on for java web start applications using myproxy.
\newblock In {\em Proceedings of the 3rd ACM workshop on Secure {W}eb
  {S}ervices}, SWS '06, pages 95--102, New York, 2006.

\bibitem{foster}
I.~Foster and C.~Kesselman.
\newblock {\em The {G}rid: Blueprint for a New Computing Infrastructure}.
\newblock Morgan Kaufmann, 1999.

\bibitem{futuregrid}
Future{G}rid: a distributed testbed exploring possibilities with clouds, grids
  and high performance computing.
\newblock \url{https://portal.futuregrid.org/}.

\bibitem{glance}
Glance, the {O}penstack {VM} image repository.
\newblock \url{http://docs.openstack.org/developer/glance/}.

\bibitem{bellecloudcomputing}
R.~Graciani~Diaz, A.~Casajus~Ramo, A.~Carmona~Ag\"{u}ero, T.~Fifield, and
  M.~Sevior.
\newblock Belle-{DIRAC} setup for using {A}mazon {E}lastic {C}ompute {C}loud.
\newblock {\em J. Grid Comput.}, 9(1):65--79, Mar. 2011.

\bibitem{gridengine}
Oracle {G}rid {E}ngine.
\newblock \url{http://www.oracle.com/}.

\bibitem{skycomputing}
K.~Keahey, M.~Tsugawa, A.~Matsunaga, and J.~Fortes.
\newblock Sky computing.
\newblock {\em Internet Computing, IEEE}, 13(5):43--51, 2009.

\bibitem{kvm}
{KVM} ({K}ernel-based {V}irtual {M}achine).
\newblock \url{http://www.linux-kvm.org/page/Main_Page}.

\bibitem{lustre}
{L}ustre {F}ilesystem.
\newblock \url{http://wiki.whamcloud.com/}.

\bibitem{nimbus}
Nimbus: an open-source {EC2}/{S3} compatible {IaaS} implementation.
\newblock \url{http://www.nimbusproject.org/}.

\bibitem{openstack}
Openstack: Open source software for building private and public clouds.
\newblock \url{http://www.openstack.org/}.

\bibitem{pbs}
{PBS} {W}orks, enabling on-demand computing.
\newblock \url{http://www.pbsworks.com/}.

\bibitem{magellan}
L.~Ramakrishnan, P.~T. Zbiegel, S.~Campbell, R.~Bradshaw, R.~S. Canon,
  S.~Coghlan, I.~Sakrejda, N.~Desai, T.~Declerck, and A.~Liu.
\newblock Magellan: experiences from a science cloud.
\newblock In {\em Proceedings of the 2nd international workshop on scientific
  cloud computing}, ScienceCloud '11, pages 49--58, New York, 2011.

\bibitem{condor}
D.~Thain, T.~Tannenbaum, and M.~Livny.
\newblock Distributed computing in practice: the {HTC}ondor experience.
\newblock {\em Concurr. Comput. : Pract. Exper.}, 17(2-4):323--356, Feb. 2005.

\bibitem{x509}
S.~Tuecke, V.~Welch, D.Engert, L.Pearlman, and M.Thompson.
\newblock Internet {X.509} public key infrastructure {(PKI)} proxy certificate
  profile.
\newblock \url{http://www.ietf.org/rfc/rfc3820.txt}.

\bibitem{repoman}
M.~Vliet, A.~Agarwal, M.~Anderson, P.~Armstrong, A.~Charbonneau, R.Demarais,
  K.~Fransham, I.~Gable, D.~Harris, R.~Impey, C.~Leavett-Brown, M.~Paterson,
  W.~Podaima, and R.~Sobie.
\newblock Repoman: A simple {REST}ful {X.509} virtual machine image repository.
\newblock {\em PoS}, ISGC 2011 and OGF 31:052, 2011.

\bibitem{webdav}
{W}eb {D}istributed {A}uthoring and {V}ersioning ({W}eb{DAV}).
\newblock \url{http://tools.ietf.org/html/rfc4918}.

\end{thebibliography}

\end{document}